\documentstyle [12pt,epsf] {article}

\def\dend#1{{\if*#1{\it Paenibacillus dendritiformis}\else
                \if-#1{\it P. dendritiformis}\else
		 {\it P. dendritiformis} #1\fi\fi}}
\def\Tvar{var. {\it dendron}}
\def\Cvar{var. {\it chiralis}}
\def\Tname#1{{\if*#1\dend* \Tvar\else
                \if-#1\dend{}\Tvar\else
                 \dend{}\Tvar{} #1\fi\fi}}
\def\Cname#1{{\if*#1\dend* \Cvar\else
                \if-#1\dend{}\Cvar\else
                 \dend{}\Cvar{} #1\fi\fi}}
\def\Vname#1{{\if*#1{\it Paenibacillus vortex}\else
                \if-#1{\it P. vortex}\else {\it P. vortex} #1\fi\fi}}
\def\bsub#1{{\if*#1{\it Bacillus subtilis}\else
                \if-#1{\it B. subtilis}\else {\it B. subtilis} #1\fi\fi}}
\def\bacil#1{\if *#1{Bacillus}\else{B.}\fi}
\def\bcirc#1{\if *#1{\it \bacil* circulans}\else
                \if -#1{\it \bacil{} circulans}\else
                   {\it \bacil{} circulans} #1\fi\fi}
\def\ecoli#1{{\if*#1{\it Escherichia coli}\else
                \if-#1{\it E. coli}\else {\it E. coli} #1\fi\fi}}
\def\salmon#1{{\if*#1{\it Salmonella typhimurium}\else
                \if-#1{\it S. typhimurium}\else
                {\it S. typhimurium} #1\fi\fi}}
\def\prot{{\it Proteus mirabilis }}
\def\myxo#1{{\if*#1{\it Myxococcus xanthus}\else
                \if-#1{\it M. xanthus}\else
                {\it M. xanthus} #1\fi\fi}}

\def\T{{${\cal T }$} }
\def\C{{${\cal C }$} }

\def\Tm{{\T morphotype }}

\def\Tme#1{{\T morphotype#1}}
\def\Cm{{\C morphotype }}

\def\Cme#1{{\C morphotype#1}}

\def\partderiv#1#2{{\partial #1\over\partial #2}}
\def\vs{{\it vs. }}
\def\etal{{\it et al. }}
\def\text#1{\hbox{#1}}

\begin{document}

\title{
{\flushleft \normalsize \tt  Proceedings of IMA workshop:\hfill \\
\flushleft Pattern Formation and Morphogenesis (1998) \hfill\\}
~\\
~\\
Modeling Branching and Chiral Colonial Patterning of
Lubricating Bacteria}
\author{{\bf
Eshel Ben-Jacob, Inon Cohen,
}\\ {\bf
Ido Golding and Yonathan Kozlovsky}
\\
School of Physics and Astronomy, \\
Raymond and Beverly Sackler Faculty of Exact Sciences, \\
Tel Aviv University, Tel Aviv 69978, Israel}
\maketitle

%-----------------------------------------------------------
% Document
%-----------------------------------------------------------

\begin{abstract}
In nature, microorganisms must often cope with hostile
environmental conditions. To do so they have developed sophisticated
cooperative behavior and intricate communication capabilities, such
as: direct cell-cell physical interactions via extra-membrane
polymers, collective production of extracellular "wetting" fluid for
movement on hard surfaces, long range chemical signaling such as
quorum sensing and chemotactic (bias of movement according to gradient
of chemical agent) signaling, collective activation and
deactivation of genes and even exchange of genetic material. Utilizing
these capabilities, the colonies develop complex
spatio-temporal patterns in response to adverse growth conditions.
We present a wealth of branching and chiral patterns formed during
colonial development of lubricating bacteria (bacteria which produce a
wetting layer of fluid for their movement).
Invoking ideas from pattern formation in
non-living systems and using ``generic'' modeling we are able to reveal
novel survival strategies which account for the salient features of
the evolved patterns. Using the models, we demonstrate how 
communication leads to self-organization via cooperative behavior of 
the cells. In this regard, pattern
formation in microorganisms can be viewed as the result of 
the exchange of information between the micro-level (the individual
cells) and the macro-level (the colony).
We mainly review known results, but include a new model of chiral
growth, which enables us to study the effect of chemotactic signaling
on the chiral growth. We also introduce a measure for weak chirality
and use this measure to compare the results of model simulations with
experimental observations.
\end{abstract}

\section{Introduction}
\label{sec:intro}

Among non-equilibrium dynamical systems, living organisms are the
most challenging ones that scientists can study.
A biological system constantly exchanges material, energy and information
with the
environment as it regulates its growth and survival.
The energy and chemical balances at the cellular level involve an
intricate interplay between the microscopic dynamics and the
macroscopic environment, through which life at the intermediate
mesoscopic scale is maintained \cite{AlbrechtB90}. The development of a
multicellular structure requires non-equilibrium dynamics, as
microscopic imbalances are translated into the macroscopic gradients
that control collective action and growth \cite{PP92}.

Much effort is devoted to the search for basic principles of
organization (growth,communication, regulation and control) on the
cellular and multicellular levels
\cite{Rosenberg84,Kessler85,Shap88,Haken88,NP89,ST91,KL93,Wiener48}.
Our approach is to use the successful conceptual framework for pattern
formation in non-living systems as a tool to unravel their
significantly more complex biological counterparts.
Of critical importance is the choice of starting point, i.e. the
choice of which phenomena to study: it has to be simple enough to
allow progress, but also well motivated by the significance of the
results.  Cooperative microbial behavior is well suited for these
requirements, as we explain below. We focus on two examples: the
branching growth in bacterial colonies of \Tname, and the chiral
growth of \Cname.  Examples of these patterns are shown in Fig.\
\ref{fig:intro}.

Traditionally, bacterial colonies are grown on substrates with a high
nutrient level and intermediate agar concentration \cite{Henrici48}.
Such ``friendly'' conditions yield colonies of simple compact
patterns, which fit well the contemporary view of bacterial colonies
as a collection of independent unicellular organisms (non-interacting
``particles''). However, bacterial colonies in nature must regularly
cope with hostile environmental conditions \cite{Shap88,SDA57}.
When  hostile conditions are created in a petri dish by using a very
low level of nutrients, a hard surface (high concentration of agar),
or both, very complex patterns are observed. 

Drawing on the analogy with diffusive patterning in non-living systems
\cite{KKL88,Langer89,BG90,BenJacob93,BenJacob97}, we can state that 
complex patterns are expected.  The cellular reproduction
rate that determines the growth rate of the colony is limited by the
level of nutrients available for the cells. The latter is limited by
the diffusion of nutrients towards the colony (for low nutrient
substrate). Hence colony growth under certain conditions should be
similar to diffusion limited growth in non-living systems as mentioned
above \cite{BG90,BenJacob93,BenJacob97}.  The study of diffusive
patterning in non-living systems teaches us that the diffusion field
drives the system towards decorated (on many length scales) irregular
fractal shapes
\cite{Mandelbrot77a,Mandelbrot77b,Sander86,Vicsek89,Feder88}. Indeed,
bacterial colonies can develop patterns reminiscent of those observed
during growth in non-living systems
\cite{FM89,MF90,FM91,MWM95,BSTA95,BSST92,BTSA94,BST94,MHM93,MM93}. 

But, this is certainly not the end of the story. In fact, the colonies
exhibit far richer behavior. This is ultimately a reflection of the
additional levels of complexity involved
\cite{BSST92,BTSA94,BSTCCV94a,BSTCCV94b,BCSALT95,BCSCV95,CBCV96,BCCVG97}.
The building blocks of the colonies are themselves living
systems, each having its own autonomous self-interest and internal
degrees of freedom. Yet, efficient adaptation of the colony to adverse
growth conditions requires self-organization on all levels which can
only be achieved via cooperative behavior of these individual cells.
Thus, pattern formation at the colony level may be viewed as the
outcome of a dynamical interplay \cite{BSST92,BTSA94,BST94} between
the micro-level (the individual cell) and the macro-level (the
colony). For this interplay to work, the effects of changes at the
micro-level must make themselves felt at the macro-level. This is why
the notion of a singular perturbation, discovered to be a key for
understanding pattern selection in non-living systems, will be of even
more importance here. 

This manuscript is mainly a review of our modeling studies. Yet we
include some important new results: 1. Explanations and modeling of
chemotaxis during chiral growth. 2. Modeling and characterization of
weak chirality (global weak twist) during branching growth.

For completeness, section \ref{sec:bio-bg} includes a brief description
of the necessary biological background needed to justify our models of
the growth.

How should one approach the modeling of the complex bacterial patterning?
With present computational power it is natural to use computer models as a
main tool in the study of complex systems. However, one must be careful not
to be trapped in the "reminiscence syndrome", described by J. D. Cowan
\cite{Horgan95}, as the tendency to devise a set of rules which will
mimic some aspect of the observed phenomena and then, to quote J. D.
Cowan "They say: `Look, isn't this reminiscent of a biological or
physical phenomenon!' They jump in right away as if it's a decent
model for the phenomenon, and usually of course it's just got some
accidental features that make it look like something." Yet the
reminiscence modeling approach has some indirect value.  True, doing
so does not reveal (directly) the biological functions and behavior.
However, it does reflect understanding of geometrical and temporal
features of the patterns, which indirectly might help in revealing
the underlying biological principles. Another extreme is the
"realistic modeling" approach, where one constructs a model that
includes in details all the known biological facts about the system.
Such an approach sets a trajectory of ever including more and more
details (vs. generalized features). The model keeps evolving to
include so many details that it loses any predictive power.

Here we try to promote another approach -- the "generic modeling" one
\cite{KL93,BSTCCV94a,Azbel93,KW97}. We seek to elicit, from the
experimental observations and the biological knowledge, the generic
features and basic principles needed to explain the biological
behavior and to include these features in the model. We will
demonstrate that such modeling, with close comparison to experimental
observations, can be used as a research tool to reveal new
understanding of the biological systems.

Generic modeling is not about using sophisticated, as it may, mathematical
description to dress pre-existing understanding of complex biological
behavior. Rather, it means a cooperative approach, using existing biological
knowledge together with mathematical tools and synergetic point of view for
complex systems to reach a new understanding (which is reflected in the
constructed model) of the observed complex phenomena.

The generic models can yet be grouped into two main categories:
1. Discrete models such as the Communicating Walkers models of
Ben-Jacob \etal \cite{BSTCCV94a,BCSCV95,BCCVG97} and the Bions model
of Kessler and Levine \cite{KL93,KLT97}.
In this approach, the microorganisms (bacteria in the first model and
amoebae in second) are represented by discrete, random walking
entities (walkers and bions, respectively) which can consume
nutrients, reproduce, perform random or biased movement, and produce
or respond to chemicals. The time evolution of the chemicals is
described by reaction-diffusion equations.
2. Continuous or reaction-diffusion models \cite{PS78,Mackay78}. In
these models the microorganisms are represented via their 2D density,
and a reaction-diffusion equation of this density describes their
time evolution. This equation is coupled to the other
reaction-diffusion equations for the chemical fields. In the context
of branching growth, this idea has been pursued recently by Mimura and
Matsushita
\etal \cite{Mimura97,MWIRMSM98},
Kawasaki \etal \cite{KMMUS97}, Kitsunezaki \cite{Kitsunezaki97} and
Kozlovsky \etal \cite{KCGB98}.
 A summary and critique of this approach can be found in
\cite{Rafols98} and \cite{GKCB98}.

In section \ref{sec:cont-models} we present the continuous modeling of
the branching growth. In section \ref{sec:C} the chiral growth is
modeled using the ``atomistic'' Communicating Spinors model, which
enables us to model chemotaxis response. It is the first time that the
chemotaxis effect on chiral growth has been studied. The actual study
of chemotaxis, both in the chiral growth and the branching growth is
done in section \ref{sec:chemo}.

Section \ref{sec:weak_chiral} is devoted to the studies of weak
chirality. The phenomenon is modeled using both continuous and
discrete models. We introduce a measure for weak chirality which
enables a more crucial comparison between the models' results and the
observed patterns. Good agreement was found.

Conclusions are presented in Section \ref{sec:conc}. We explain that
the weak chirality phenomenon is general, and show examples of weak
chirality during growth of the chiral morphotype and the vortex
morphotype. In the latter it results from a different mechanism, and
indeed the twist is not linear with the radius of growth.

\section{Observations and Biological background}
\label{sec:bio-bg}
Following the experimental observations which are explained in this
manuscript, we will describe the most relevant information for the
understanding and modeling of the observed colonial patterning. We
base relevancy on our previous experience and we concentrate on
bacterial movement.

\subsection{Experimental observations: branching growth of bacterial
colonies}
\label{sec:T}

\subsubsection{Macroscopic Observations}

Some additional examples of the patterns exhibited by colonies of the
\Tm are shown in figures \ref{fig:2.1}, \ref{fig:2.2} and
\ref{fig:2.3}. For intermediate agar concentrations (about 1.5\% --
$1.5g$ in $100ml$), at very high peptone levels (above 10$g/l$) the
patterns are compact (Fig. \ref{fig:2.2}a).  At somewhat lower but
still high peptone levels (about 5-10$g/l$) the patterns exhibit
quite pronounced radial symmetry and may be characterized as dense
fingers (Fig. \ref{fig:2.2}b), each finger being much wider than the
distance between fingers.
For intermediate peptone levels, branching patterns with lower
fractal dimension (reminiscent of electro-chemical deposition) are
observed (Fig. \ref{fig:2.2}c). The patterns are ``bushy'', with
branch width smaller than the distance between branches.
As the peptone level is lowered, the patterns become more ramified
and fractal--like. Surprisingly, at even lower peptone
levels (below 0.25$g/l$ for 2\% agar concentration) the colonies
revert to organized structures: fine branches forming a well
defined global envelope. We characterize these patterns as fine
radial branches (Fig. \ref{fig:2.2}d). For extremely low peptone
levels (below 0.1$g/l$), the colonies lose the fine radial structure
and again exhibit fractal patterns (Fig. \ref{fig:2.3}).  For high
agar concentration the branches are very thin (Fig. \ref{fig:2.3}b).

At high agar concentration and very high peptone levels the colonies
display a structure of concentric rings (Fig. \ref{fig:2.4}). At high
agar concentrations the branches also exhibit a global twist with the
same handedness (weak chirality), as shown in Fig. \ref{fig:2.5}.
Similar observations during growth of other bacterial strains have
been reported by Matsuyama \etal \cite{MM93,MHM93}. We referred to
such growth patterns as having weak chirality, as opposed to the
strong chirality exhibited by the \Cme.

A closer look at an individual branch (Fig.  \ref{fig:2.6}) reveals a
phenomenon of density variations within the branches. These
3-dimensional structures arise from accumulation of cells in layers.
The aggregates can form spots and ridges which are either scattered
randomly, ordered in rows, or organized in a leaf-veins-like structure.
The aggregates are not frozen; the cells in them are motile and the
aggregates are dynamically maintained.

\subsubsection{Microscopic Observations}
\label{sec:micro}

Under the microscope, cells are seen to perform a random-walk-like
movement in a fluid. This fluid is, we assume, excreted by the cells
and/or drawn by the cells from the agar \cite{BSTCCV94a,BSTCCV94b}.
The cellular movement is confined to this fluid; isolated cells
spotted on the agar surface do not move. The boundary of the fluid
thus defines a local boundary for the branch (Fig.  \ref{fig:2.9}).
Whenever the cells are active, the boundary propagates slowly as a
result of the cellular movement pushing the envelope forward and
production of additional wetting fluid. Electron microscope
observations reveal that these bacteria have flagella for swimming.

The observations reveal also that the cells are active at the outer parts of 
the colony, while closer to the center the cells are stationary and some of 
them sporulate (form spores) (Fig.  \ref{fig:2.10}). It is known that
certain bacteria respond to adverse growth conditions by entering a
spore stage until more favorable growth conditions return. Such spores
are metabolically inert and exhibit a marked resistance to the lethal
effects of heat, drying, freezing, deleterious chemicals, and
radiation.

At very low agar concentrations (below 0.5\%) the bacteria swim
inside the agar and not on its surface. Between 0.5\% and 1\% agar
concentration some of the bacteria move on the surface and some inside
the agar.

\subsection{Chiral patterns}

Chiral asymmetry (first discovered by Louis Pasteur) exists
in a whole range of scales, from subatomic particles through human
beings to galaxies, and seems to have played an important role in the
evolution of living systems \cite{HK90,AGK91}.
Bacteria display various chiral properties. Mendelson \etal
\cite{Mend78,MK82,MT89,Mend90} showed that long cells of \bsub can grow
in helices, in which the cells form long strings that twist around
each other. They have shown also that the chiral characteristics
affect the structure of the colony. Ben-Jacob \etal 
\cite{BSST92,BTSA94,BCSCV95} have found yet another chiral
property -- the strong chirality exhibited by the \Cme. Here, 
the flagella handedness acts as a microscopic perturbation which
is amplified by the diffusive instability, leading to
the observed macroscopic chirality. This appears to be analogous to
the manner in which crystalline anisotropy leads to the observed symmetry of 
snowflakes
\cite{BenJacob93}; more about this later.

\subsubsection{A Closer Look at the Patterns}

\Cm exhibits a wealth of different patterns according to the growth
conditions (Fig. \ref{fig:2.14}).
As for \Tme, the patterns are  generally compact at high
peptone
levels and become ramified (fractal) at low peptone levels. At very
high peptone levels and high agar concentration, \Cm conceals its
chiral nature and exhibits branching growth similar to that of \Tme.

Below 0.5\% agar concentration the \Cm exhibits compact growth with
density variations. These patterns are almost
indistinguishable from those developed by the \Tme.
In the range of 0.4\%-0.6\% agar concentration the \Cm exhibits its most
complex patterns (Fig. \ref{fig:2.16}).
Surprisingly, these patterns are composed of chiral branches of both
left and right handedness.
Microscopic observations reveal that part of the growth is on top of
the agar surface while in other parts the growth is in the agar. Our
model of the chiral growth explains that indeed growth on top of the
surface and in the agar should lead to opposite handedness.

Optical microscope observations indicate that during growth of
strong chirality the cells move within a well defined envelope. The
cells are long relative to those of \Tme, and the movement appears
correlated in orientation (Fig. \ref{fig:2.17}).  Each branch tip
maintains its shape, and at the same time the tips keep twisting with
specific handedness while propagating. Electron microscope
observations do not reveal any chiral structure on the cellular
membrane \cite{BSTCCV94b}.

\subsection{Biological background}

\subsubsection{Bacterial movement}

In the course of evolution, bacteria have developed ingenious
ways of moving on surfaces. The
most widely studied and perhaps the most sophisticated translocation
mechanism used by bacteria is the flagellum \cite{Eis90}, but other
mechanisms exist as well \cite{Henri72}.
Swimming  is a solitary movement done in liquid. A swimming bacterium
runs nearly straight runs, interrupted by short periods of tumbling.
Tumbling event is a random rotation in one location.
The direction of the next run is dictated by the final orientation of
the tumbling bacterium.

A swimming bacterium propagates itself by rotating a bundle of
flagella.  Each flagellum is an helical protein filament which is
hooked to a molecular engine transversing the bacterial membrane. The
engines of all flagella rotate synchronously clockwise or
counterclockwise. When the bacterium turn them counterclockwise, the
flagella form an aligned bundle and push the bacterium forward. When
they turn clockwise, the flagella disjoin and the bacterium tumbles.

In our experiments, the bacteria do not swim in a pre-supplied liquid,
but in a layer of fluid on the surface of the agar.
The bacterial cells move individually and at random in the same manner
as flagellated bacteria move in wet mounts (i.e., nearly straight runs
separated by brief tumbling). Swimming takes place only in
sufficiently thick surface fluid. Microscope observations reveal no
organized flow-field pattern.

Based on microscope observations of movement and electron microscope
observations of flagella we identify the movement of \C and \Tm as
swimming.
Cells tumble about every $\tau_T\approx 1-5~sec$ depending on external
conditions. The speed of the bacterium between tumbling events is very
sensitive to conditions such as the liquid viscosity, temperature and
pH level. Typically, it is of the order of 1-10$\mu
m/sec$.

Swimming can be approximated by a random walk with variable step size
\cite{Berg93}.
At low bacterial densities the random walk can be described by a
diffusion equation with a
diffusion coefficient $D_b\equiv v^2\,\tau
_T=10^{-8}-10^{-5}cm^2/sec$.
Low bacterial densities means that the mean free path between
bacterial collisions $l_c$ is longer than the tumbling length
$l_T\equiv v\tau _T $, thus
collisions between the bacteria can be neglected. The mean free path
(or collision length) is
\begin{equation}
l_c\propto \left\{
\begin{array}{ll}
\rho ^{-\frac{{1}}3} & \hbox{in 3 dimensions} \\
\sigma ^{-\frac{{1}}2} & \hbox{in 2 dimensions}
\end{array}
\right.
\end{equation}
where $\rho $ is the 3D bacterial density and $\sigma $ is the 2D
density -- the projection of $\rho $ on the surface.

At high densities ($l_c<l_T$), the collisions cannot be
neglected.  In attempt to approximate the dynamics in those
conditions, one may want to consider the time of straight motion to be
$l_c/v$ instead of $\tau _T$.
Hence $D_b$ depends on the bacterial density to yield
\begin{equation}
D_b\propto \left\{
\begin{array}{ll}
v\rho ^{-\frac{{1}}3} & \hbox{in 3D} \\
v\sigma ^{-\frac{{1}}2} & \hbox{in 2D}
\end{array}
\right. ~~.  \label{eq:hige_density_diffusion:0}
\end{equation}
This approximation is valid under the assumptions that a collision
event is identical to a tumbling event (abrupt uncorrelated change in
direction of motion), that a tumbling event is independent of the
collisions, and that the speed between such events is not affected by
their frequency.

The assumption that a collision event is like a tumbling event poses
many problems. Even if the bacteria do not activate special response
to collision it is unrealistic to assume that collisions are elastic,
or that the flagella adopt immediately to the new orientation which
changes during collisions.  Thus it
is reasonable to assume strong correlation between the cell's
orientation before collision and the cell's orientation after
collision. In addition, the orientation after the collision should be
biased according with the average direction of motion of the
surrounding bacteria, as they carry the liquid with them.
The important parameter is not the collision length $l_c$
but re-orientation time $\tau_r$. The re-orientation time is the time
it takes a bacterium to loose memory of its initial orientation, i.e.
the time span on which the final orientation has effectively no
correlation with the initial orientation. At low densities the
re-orientation time $\tau_r$ is equal to the tumbling time $\tau_T$.
As the density rises and the collisions become more frequent, $\tau_r$
decrease.  $ \tau_r$ defines the densities above which the constant
diffusion coefficient $D_b\equiv v^2\,\tau _T$ is not a good
approximation. It is quite possible that these densities are high
enough so as to make the velocity and even the type of motion
dependent on bacterial density, making relation (\ref
{eq:hige_density_diffusion:0}) irrelevant. In any case, high cellular
densities does mean an effective decrease in the diffusion coefficient
related to the bacterial movement.

When swimming in an unstirred liquid, very low cellular densities
also effect the movement. The bacteria secrete various materials into
the media and some of them, e.g. enzymes and other polymers,
significantly change the physical properties of the liquid making it
more suitable for bacterial swimming. The secretion of these materials
depend on cellular density, thus at not-too-high densities the speed
of swimming rise with the cellular density. Hence the diffusion
coefficient related to the bacterial movement should be a
non-monotonic function of the bacterial density.  Moreover, the
specific functional form might depend on the specific bacterial
strain.

In other conditions there is similar but more pronounced effect.
On semi-solid surface the bacteria cannot swim at all inside the agar
and they have to produce their own layer of liquid to swim in it.
To produce such fluid the bacteria secrete lubricant (wetting agents).
Other bacterial species produce known extracellular lubricants
(such as surfactants, see \cite{MKNIHY92,MNZ93,DB97,MN96} and
references therein, or the extracellular slime produced by \prot
\cite{SSW83}).
These are various materials (various cyclic lipopeptides were
identified) which draw water from the agar.  The composition and
properties of the lubricant of \dend is not known, but we will assume
that higher concentration of lubricant is needed to extract water from
a dryer agar, and that the lubricant is slowly absorbed into the agar
(or decomposes).
A single bacterium on the agar surface cannot produce enough fluid
to swim in it, thus the bacteria cannot break out of the layer fluid
and the branches of a \T or \C colony can be defined by this fluid.
Whenever bacteria enter the shallower parts of the layer, at the edge
of the branch, they become sluggish, indicating that the depth of the
layer effects the bacterial movement.
It can be argued (see section \ref{nonlinear}) that in such cases the
bacterial speed is related to the bacterial density by a power law (at
least in low densities). Not only the diffusion coefficient related to
the bacterial movement is a non-monotonic function of the bacterial
density (as in a liquid agar), but it is also vanishes for extremely
low densities.  In this case it is clear that the specific functional
form depend on the specific bacterial strain (\bsub-, for example,
cannot move at all under such conditions).

\subsubsection{Chemotaxis in swimming bacteria}

\label{sec:bg:chemotaxis}

Chemotaxis means changes in the movement of the bacteria in response
to a gradient of certain chemical field
\cite{Adler69,BP77,Lacki81,Berg93}.  The movement is biased along the
gradient either in the gradient direction or in the opposite
direction. Usually chemotactic response means a response to an
externally produced field, like in the case of chemotaxis towards
food.
However, the chemotactic response can be also to a field produced
directly or indirectly by the bacterial cells. We will refer to this
case as chemotactic signaling. The bacteria sense the local
concentration $R$ of a chemical by membrane receptors binding the
chemical's molecules \cite{Adler69,Lacki81}. It is crucial to note
that when estimating gradients of chemicals, the bacterial cells
actually measure changes in the receptors' occupancy and not in the
concentration itself. When put in continuous equations
\cite{Murray89,GKCB98}, this indirect measurement translates to
measuring the gradient
\begin{equation}
\frac \partial {\partial x}\frac R{\left( K+R\right) }=\frac
K{(K+R)^2}\frac{\partial R}{\partial x}.
\label{eq:bioBG:receptorLow}
\end{equation}
where $K$ is a constant whose value depends on the receptors'
affinity, the speed in which the bacterium processes the signal from
the receptor, etc.
This means that the chemical gradient times a factor $K/(K+R)^2$ is
measured, and it is known as the ``receptor law'' \cite{Murray89}.

In a continuous model, we incorporate the effect of chemotaxis by
introducing a chemotactic flux $ \vec{J}_{chem}$:
\begin{equation}
\vec{J}_{chem}\equiv \zeta (\sigma )\chi (R)\nabla R
\label{j_chem}
\end{equation}
$\chi (R)\nabla R$ is the gradient sensed by the bacteria (with $\chi
(R)$ having the units of 1 over chemical's concentration). $\chi (R)$
is usually taken to be either constant or the ``receptor law''. $\zeta
(\sigma )$ is the bacterial response to the sensed gradient (having the
same units as a diffusion coefficient times the units of the bacterial
density $\sigma$). It is positive for attractive
chemotaxis and negative for repulsive chemotaxis.

Ben-Jacob \etal argued \cite{BCC96,CCB96,BC97,BenJacob97} that for
the colonial adaptive self-organization the bacteria employ three
kinds of chemotactic responses, each dominant in different regime of
the morphology diagram (the claim was made for \Tme, but the same hold
for their relatives \Cme). One response is the food chemotaxis
mentioned above. It is expected to be dominant for only a range of
nutrient levels (see the ``receptor law'' below). The two other kinds
of chemotactic responses are signaling chemotaxis. One is long-range
repulsive chemotaxis. The repelling chemical is secreted by starved
bacteria at the inner parts of the colony. The second signal is a
short-range attractant. The length scale of each signal is determined
by the diffusion constant of the chemical agent and the rate of its
spontaneous decomposition.

{\em Amplification of diffusive Instability Due to Nutrients
Chemotaxis:}
In non-living systems, more ramified patterns (lower fractal
dimension) are observed for lower growth velocity. Based on growth
velocity as function of nutrient level and based on growth dynamics,
Ben-Jacob \etal \cite {BSTCCV94a} concluded that in the case of
bacterial colonies there is a need for mechanism that can both
increase the growth velocity and maintain, or even decrease, the
fractal dimension. They suggested food chemotaxis to be the required
mechanism. It provides an outward drift to the cellular movements;
thus, it should increase the rate of envelope propagation.  At the
same time, being a response to an external field it should also
amplify the basic diffusion instability of the nutrient field. Hence,
it can support faster growth velocity together with a ramified pattern
of low fractal dimension.

{\em Repulsive chemotactic signaling: }
\label{sec:repulsive}
We focus now on the formation of the fine radial branching patterns at
low nutrient levels. From the study of non-living systems, it is
known that in the same manner that an external diffusion field leads
to the diffusion instability, an internal diffusion field will
stabilize the growth. It is natural to assume that some sort of
chemotactic agent produces such a field. To regulate the organization
of the branches, it must be a long-range signal. To result in radial
branches it must be a repulsive chemical produced by bacteria at the
inner parts of the colony. The most probable candidates are the
bacteria entering a pre-spore stage.

If nutrient is deficient for a long enough time, bacterial cells may
enter a special stationary state -- a state of a spore -- which
enables them to survive much longer without food. While the spores
themselves do not emit any chemicals (as they have no metabolism), the
pre-spores (sporulating cells) do not move and emit a very wide range
of waste materials, some of which unique to the sporulating cell.
These emitted chemicals might be used by other bacteria as a signal
carrying information about the conditions at the location of the
pre-spores. Ben-Jacob \etal \cite{BSTCCV94a,BSTCCV94b,CCB96} suggested
that such materials are repelling the bacteria ('repulsive chemotactic
signaling') as if they escape a dangerous location.

\subsubsection{Food Consumption, Reproduction and Starvation}

\label{sec:eating}

\dend-, like most bacteria, reproduce by fission of
the cell into two daughter cells which are practically identical to
the mother cell. The crucial step in the cell division is the
replication of the genetic material and its sharing between the
daughter cells. Haste replication of DNA might lead to many errors --
most organisms limit the rate of replication to about 1000 bases per
second.  Thus the reproduction must take at least minimal reproduction
time $\tau _R$.  This reproduction time $\tau _R$ is about $25min$ in
Bacilli.

For reproduction, as well as for movement and other metabolic
processes, bacteria and all other organisms need influx of energy. Any
organism which does not get its energy directly from sunlight (by
photo-synthesis) needs an external supply of food. In the patterning
experiments the bacteria eat nutrient from the agar. As long as there
is enough nutrient and no significant amount of toxic materials, food
is consumed (for cell replication and internal processes) at maximal
rate $\Omega _c$. To estimate $\Omega _c$ we assume that a bacterium
needs to consume an amount of food $ C_R$ of about $3\times
10^{-12}g$.  It is 3 times its weight -- one quanta for doubling body
mass, one quanta used for movement and all other metabolic processes
during the reproduction time $ \tau _R$, and one quanta is for the
reduced entropy of making organized cell out of food.  Hence $\Omega
_c$ is about 2$fg/sec$ (1$fg=10^{-15}$gram).

If nutrient is deficient for a long enough period of time, the
bacterial cells may enter a special stationary state -- a state of
a spore -- which enables them to survive much longer without food.
The bacterial cells employ very complex mechanisms tailored for the
process of sporulation.
They stop normal activity -- like movement -- and use all their
internal reserves to metamorphose from an active volatile cell to a
sedentary durable 'seed'. While the spores themselves do not emit any
chemicals (as they have no metabolism), the pre-spores (sporulating
cells, see Fig. \ref{fig:2.10}) do not move and emit a very wide
range of wast e materials, some of which unique to the sporulating
cell. These emitted chemicals might be used by other cells as a signal
carrying information about the conditions at the location of the
pre-spores. Ben-Jacob {\it et al.} \cite{BSTCCV94a,BSTCCV94b,CCB96}
suggested that such materials are repelling the bacteria ('repulsive
chemotactic signaling') as if they escape a dangerous location.

When bacteria are grown in a petri dish, nutrients are usually
provided by adding peptone, a mixture including all the amino acids
and sugars as source of carbon. Bacteria which are not
defective in synthesis of any amino acid can grow also on a
minimal agar in which a single source of carbon and no amino acids
are provided.
Such growth might seem to be easier to model as the growth is limited
by the diffusion of a single chemical. However, during growth on
minimal agar there is usually a higher rate of waste products
accumulation, introducing other complications into the model.
Moreover many of our strains are auxotrophic i.e. defective in
synthesis of some amino acids and need an external supply of it.
Providing the bacteria with these amino acids and only a  single
carbon source might pose us the question as to what is the limiting
factor in the growth of the bacteria.  For all those reasons we prefer
to use peptone as nutrient source.

We said that if there is ample supply of food, bacteria reproduce in a
maximal rate of one division in $\tau_R$. If the available amount of
food is limited, bacteria consume the maximum amount of food they can.
In the limit of low bacterial density, the available amount of food
over the tumbling time $\tau _T$ is the food contained in the area
$\tau _T\sqrt{D_bD_n}$, where $D_b$ and
$D_n$ are the diffusion coefficients of the bacteria and the food,
respectively. Hence the rate of food consumption is given by
$n\sqrt{D_bD_n}$ (weather $D_b$ is constant or not).

In a continuous model, reproduction of bacteria translate to a growth
term of the bacterial density which is $\sigma $ times the eating rate
per bacteria. In the limit of high nutrient it is
$\sigma /\tau_R$, and in the limit of low nutrient it
is proportional to $n\sigma $. This brings to mind Michaelis-Menten
law \cite{Murray89} of $\frac K{1+\gamma n}n\sigma $ with $K,\gamma $
constants. Many authors take only the low nutrient limit of this
expression, $K n\sigma$, although it is not biologically established
that the bacteria in the experiments are limited by the availability
of food and not by their maximal consumption rate.

\section{Continuous models for the branching growth of \T morphotype}
\label{sec:cont-models}
\subsection{The Lubricating Bacteria Model}
\label{sec:lubrication}
The Lubricating Bacteria model is a reaction-diffusion model for the bacterial 
colonies of the \Tm \cite{GKCB98,KCGB98}. This model includes four
coupled fields. 
One field describes the bacterial density $b(\vec{x},t)$, the second describes 
the height of lubrication layer in which the bacteria swim $l(\vec{x },t)$, a 
third field describes the nutrients $n(\vec{x},t)$ and the fourth field is the 
stationary bacteria that ``freeze'' and begin to sporulate $s( \vec{x},t)$ (see 
section \ref{sec:repulsive}). 

We first describe the dynamics of the bacteria and of the nutrient.
The two reaction-diffusion equations governing those fields are:
\begin{eqnarray}
  \label{eq-1}
  \partderiv{b}{t}&=&movement+ \Gamma_b(b,n)
  \nonumber\\
  \partderiv{n}{t}&=&D_n\nabla^2 n - g(n,b)
\end{eqnarray}
where $\Gamma_b(b,n)$ is the bacterial reproduction term.
The nutrient diffusion is a simple diffusion process with a constant 
diffusion coefficient $D_n$.
The bacteria consume nutrients at the rate $g(n,b)$ which is taken to be:
\begin{equation}
\label{eqn-g}
  g(n,b)= nb
\end{equation}
This approximate term is correct at the limit of low nutrient level 
and low bacterial density.

The nutrient consumed by bacteria serves as an energy source 
and as a precursor for synthesis of macromolecules.
There is probably a minimum amount of energy necessary to maintain
cell structure and integrity, called maintenance energy, and nutrients
used to supply the maintenance energy are not available for cell growth.
We assume that those nutrients are required at a constant 
rate $\mu$.
Bacteria then cannot utilize all the nutrients for reproduction:
\begin{equation}
\label{eq-gamma}
\Gamma_b(b,n) =  g(n,b) - \mu b 
\end{equation}

There is no explicit term for sporulation in (\ref{eq-1}). 
Instead the reproduction term $\Gamma_b$ can be used also to model sporulation.
Sporulation is initiated by starvation so it is complementary to reproduction.
When $\Gamma_b>0$ bacteria reproduce and so sporulation is excluded.
In the other case when $\Gamma_b<0$ bacteria reduce in number.
This should represent sporulation.
Note that $\Gamma_b<0$ when the nutrient level is low so indeed 
we can claim that bacteria are starved.
This simple sporulation scheme has its limitations. Its rate
is set by the nutrient consumption and effects such as density
dependence are neglected.
We represent the sporulating bacteria by the field $s(\vec{x},t)$, 
whose time evolution is:
\begin{equation}
\label{eq-s}
\partderiv{s}{t} = \left\{ \begin{array}{ll}
0 & \mbox{~~~if } \Gamma_b > 0 \\
-\Gamma_b & \mbox{~~~if } \Gamma_b < 0 \end{array} 
\right.
\end{equation}

In other continuous models the bacterial reproduction and the
sporulation process are modeled differently
\cite{Kitsunezaki97,KMMUS97,MWIRMSM98,GKCB98}.
The bacterial reproduction is proportional to
the nutrient consumption rate and sporulation is an independent
process that proceeds at a rate $\mu$ which could depend on other
variables such as the bacterial density or the nutrient concentration.
Note that if those modifications are applied to this model, 
the functional form of the bacterial reaction terms 
(\ref{eq-1}), (\ref{eq-gamma}) will not change.
The equation for the field $s$ (\ref{eq-s}) will change to:
\begin{equation}
\partderiv{s}{t} = \mu b
\end{equation}  
Since the dynamics of the other variables are separated from $s$,
its different dynamics are not significant.
Moreover, the modification of the equation for $s$ is minor.
The difference is only in the biological interpretation of the terms.

We now turn to the bacterial movement. In a uniform layer of liquid, 
bacterial swimming is a random walk with a variable step length and can 
be approximated by diffusion. The layer of lubricant is not 
uniform, and its height affects the bacterial movement. An increase 
in the amount of lubricant decreases the friction between the 
bacteria and the agar surface. 
We suggest that the bacterial movement depends on the local lubricant 
height through a power law with the exponent $\gamma>0$:
\begin{equation}
  \label{eq-bactdif}
  movement=\nabla \cdot (D_b l^\gamma \nabla b)
\end{equation}
where $D_b$ is a constant with dimensions of a diffusion coefficient.
$D_b$ is related to the fluid's viscosity and the dryness of the agar
might affect this viscosity. 
Gathering the various terms gives the partial model:
\begin{eqnarray} 
  \label{eq3}
  \partderiv{b}{t}&=&\nabla \cdot (D_b l^\gamma \nabla b)+ nb-\mu b
  \nonumber\\
  \partderiv{n}{t}&=&D_n\nabla^2n-nb \nonumber \\
  \partderiv{s}{t}&=&-\min \left( nb-\mu b,0\right)  
\end{eqnarray}
It is possible to define dimensionless time and space variables
$t'=t\mu$ and $\vec{x'}=\vec{x}\sqrt{\mu/D_n}$. In those units the
parameters $D_n$ and $\mu$ are equal to $1$.

We model the dynamics of the lubricating fluid also by a reaction 
diffusion equation. There are two reaction terms: production by the 
bacteria and absorption into the agar.
The dynamics of the field are:
\begin{equation}
  \label{eql-words}
  \partderiv{l}{t}=-\nabla  \vec{J_l} + f_l(b,n,l) - \lambda l
\end{equation}
where $\vec{J_l}$ is the fluid flux, 
$f_l(b,n,l)$ is the fluid production term and 
$\lambda$ is the absorption rate of the fluid into the agar. 

We assume that the fluid production depends on the bacterial density.
As the production of lubricant probably demands substantial metabolic 
efforts, it should also depend on the nutrient's level. We take a
simple form where the production depends linearly on the concentrations 
of both the bacteria and the nutrient.
The exact relation should depend on the synthetic pathway
of the active agents composing the lubricant. 
If they are, for example, secondary metabolites,
then their production does not depend on the current nutrient level, 
but on the prior accumulation of primary metabolites. 
However, the model is not sensitive to the exact dependence of 
the lubricant production on the
nutrient's level since the lubricant production is important at the 
front of the expanding colony, where the nutrient level is close
to the initial level.
It is reasonable that the bacteria produce lubricant up to a height,
denoted as $l_M$, which is sufficient for their swimming motion.
We therefore take the production term to be:
\begin{equation}
  f_l(b,n,l)=\Gamma bn(l_M-l)
\end{equation}
where $\Gamma$ is the production rate.

We turn to the flow of the lubricating fluid. 
The physical problem is very complicated.
However a simplified model will suffice.
We model the lubricant flux as a non-linear diffusion process: 
\begin{equation}
\label{j-l}
  \vec{J_l}=-D_l l^\nu \nabla l
\end{equation}
where $D_l$ is a constant with dimensions of a diffusion coefficient.
The diffusion term of the fluid depends on the height of 
the fluid to the power $ \nu >0$. The nonlinearity causes the fluid to 
have a sharp boundary at the front of the colony, as is observed in 
bacterial colonies.
The equation for the lubricant field is:
\begin{equation} 
  \label{eql}
  \partderiv{l}{t}=\nabla \cdot (D_l l^\nu \nabla l)
                   +\Gamma bn(l_M-l) - \lambda l 
\end{equation}

The functional form of the terms that we introduced are simple and plausible,
but they are not derived from basic physical principles. 
Therefore we do not have quantitative relations between the parameters of 
those terms and the physical properties of the agar substrate. 
However we can suggest some relations.
In the experiments, the agar concentration is controlled. 
Higher agar concentration gives a drier and more solid substrate.
We shall try to find what are the effects on the lubricant layer.
We recall that the lubricant fluid is composed from water
and active components such as surfactants.
A drier agar can increase the absorption rate $\lambda$.
Alternatively it can diminish the amount of water extracted by the
active components. Then either the lubricant layer will be thinner
or the bacteria will have to produce more of the active components.
The former case should decrease $D_b$ while the latter should decrease
the production rate $\Gamma$.
In both cases the composition of the lubricant fluid will change 
as the concentration of the active components will increase.
The lubricant fluid should become more viscous, with the effect
of $D_b$ and $D_l$ decreasing.

Equation (\ref{eql}) together with equations (\ref{eq3}) form the
Lubricating Bacteria model.
For the initial conditions, we set $n$ to have a uniform distribution 
of level $n_0$, $b$ to be zero everywhere but in the center, and the 
other fields to be zero everywhere.

Our results show that the model can reproduce branching patterns, similar to 
the bacterial colonies.
In the experiments there are two control parameters:
the agar concentration and the initial nutrient concentration.
First we examine the effect of changing the initial nutrient 
concentration $n_0$. 
As Fig. \ref{fig-nutrient} shows, 
the model produced a dense circular colony when $n_0$ was large.
The pattern became more branched and ramified as
$n_0$ decreased until $n_0$ was close to $1$, 
which is the minimal value of $n_0$ to support growth. 

Changing the agar concentration affects the dynamics of the lubricant
fluid. Previously we demonstrated that a higher agar concentration
relates to a larger absorption rate $\lambda$ and to lower production
rate $\Gamma$ and lower diffusion coefficients $D_l$ and $D_b$.  In
Fig. \ref{fig-lambda} we show patterns obtained with different values
of the parameters $\Gamma$ and $\lambda$. 
As we expected, increasing $\lambda$ or decreasing $\Gamma$ produced a
more ramified pattern, similar to the effect of a higher agar
concentration on the patterns of bacterial colonies. Similar effects
are obtained by decreasing $D_b$.

\subsection{The Non-Linear Diffusion Model}
\label{nonlinear}

Under certain assumptions, the Lubricating Bacteria model can be
reduced to the non-linear diffusion model of Kitsunezaki
\cite{Kitsunezaki97} and Cohen \cite{Cohen97}. All the additional
assumptions needed are about the dynamics near zero bacterial density:
\\
 1) The lubricant height $l$ is much smaller than $l_{max}$, so that
the production of the lubricant can be assumed to be independent of
its height \\ 
 2) The production of lubricant is proportional to the bacterial
density to the power $ \alpha>0$ (in the simplest case taken above
$\alpha=1$)\\
 3) The absorption of the lubricant is proportional to the lubricant
height to the power $ \beta>0$ (in the simplest case taken above
$\beta=1$).\\ 
 4) Over the bacterial length scale, the two above processes are much
faster than the diffusion process, so the lubricant height is
proportional to the bacterial density to the power of $
\beta/\alpha$.\\
 5) The friction is proportional to the lubricant height to the power
$ \gamma<0$.\\
Given these assumptions, the lubricant field can be removed from the 
dynamics and be replaced by a density dependent diffusion coefficient.
This diffusion coefficient is proportional to the bacterial density to
the power $k \equiv -2 \gamma \beta/\alpha > 0$.

The resulting model is:
 \begin{eqnarray}
\label{kitsunezaki-eqn}
\lefteqn{ \partderiv{b}{t} = \nabla (D_0 b^k \nabla
 b) + n b - \mu b}\\
\lefteqn{ \partderiv{n}{t} = \nabla ^2 n - b n}\\
\lefteqn{ \partderiv{s}{t} = \mu b}
\end{eqnarray}
For $k>0$ the 1D model gives rise to a front ``wall'', with compact
support (i.e. $b=0$ outside a finite domain). For $k>1$ this wall has
an infinite slope. 
The model exhibits branching patterns for suitable parameter values
and initial conditions, as depicted in Fig. \ref{fig:k1}. Increasing
initial levels of nutrient leads to denser colonies, similar to the
observed patterns.

\section{The Communicating Spinors Model for the chiral growth of \C
morphotype}
\label{sec:C}

The Communicating Spinors Model was developed to explain the chirality
of the \Cm colonies.   Our purpose is to show that the flagella handedness, 
while acting as a singular perturbation, leads to the observed chirality. 
It does so in the same manner in which crystalline anisotropy leads to 
the observed symmetry of snowflakes \cite{BenJacob93}.

It is known \cite{Eis90,SSM90,Shaw91} that flagella have specific 
handedness. Ben-Jacob \etal \cite{BCSCV95} proposed that the latter 
is the origin of the observed chirality. In a fluid (which is the state in 
most experimental setups), as the flagella unfold, the cell tumbles and 
ends up at a new random angle relative to the original one. The 
situation changes for quasi 2D motion -- motion in a "lubrication" layer 
thinner then the cellular length. We assume that in this case, of rotation 
in a plane, the tumbling has a well defined handedness of rotation. 
Such handedness requires, along with the chirality of the flagella, the 
cells' ability to distinguish up from down. The growth in an upside-
down petri- dish shows the same chirality. Therefore, we think that the 
determination of up \vs down is done either via the vertical gradient of 
the nutrient concentration or via the vertical gradient of signaling 
materials inside the substrate or via the friction of the cells with the 
surface of the agar. The latter is the most probable alternative; soft 
enough agar enables the bacteria to swim {\it below} the surface of the 
agar which leads to many changes in the patterns, including reversing 
the bias of the branches.

To cause the chirality observed on semi-solid agar, the rotation of 
tumbling must be, on average, less than $90^\circ$ and relative to a 
specific direction. Co-alignment (orientational interaction) limits the 
rotation. We further assume that the rotation is relative to the local 
mean orientation of the surrounding cells.

To test the above idea, we included the additional assumed features in 
the Communicating Walkers model \cite{BSTCCV94a}, changing it to 
a `Communicating Spinors' model (as the particles in the new model 
have an orientation and move in quasi-1D random walk). The 
Communicating Walkers model \cite{BSTCCV94a} was inspired by 
the diffusion-transition scheme used to study solidification from 
supersaturated solutions \cite{SKBLM92a,SKBLM92b,Shochet95}. 
The former is a hybridization of the ``continuous'' and ``atomistic'' 
approaches used in the study of non-living systems. Ben-Jacob \etal 
have presented in the past a version of the Communicating Spinors 
model for the chiral growth \cite{BCSCV95}. The model we present 
here is closely related to a previous model of the chiral growth, but it 
differs in two features. The first is the orientation field (see below), 
which was discontinuous piecewise constant and in this model it is 
continuous piecewise linear. The second difference is the definition of a 
single run of a spinor (the stretch between two tumbling events), which 
was defined as one run per one time unit (i.e. each step is a run) and 
now is defined as variable number of steps in the same direction.

The representation of bacteria as spinors allows for a close relation to 
the bacterial properties. The bacterial cells are represented by spinors 
allowing a more detailed description. At the end of the growth of a 
typical experiment there are $10^8-10^{9}$ bacterial cells in the petri-
dish. Thus it is impractical to incorporate into the model each and 
every cell. Instead, each of the spinors represents about $10-1000$ 
cells, so that we work with $10^4-10^6$ spinors in one numerical 
``experiment''.

Each spinor has a position $ \vec{r}_i$, direction $ \theta_i$ (an 
angle) and a metabolic state ('internal energy') $ E_i$. The spinors 
perform an off-lattice constrained random walk on a plane within an 
envelope representing the boundary of the wetting fluid. This envelope 
is defined on the same tridiagonal lattice where the diffusion equations 
are solved. To incorporate the swimming of the bacteria into the 
model, at each time step each of the active spinors (motile and 
metabolizing, as described below) recalculate its direction $ \theta'_i$ 
and moves a step of size $d$ to this direction.

The direction in which each spinor moves is determined in two steps; 
first the spinor decides whether it should continue the current run, that 
is to continue in the same direction $ \theta'_i = \theta_i$. In the basic 
version of the model (see Sec. \ref{sec:Cchemo} for extension of the 
model) the decision is random with a specific probability $p$ to 
continue the run. The resulting runs have geometric distribution of 
lengths, with mean run length of $d/p$. Once a spinor decides to 
change direction, the new direction $ \theta'_i$ is derived from the 
spinor's previous direction by
\begin{equation}
\theta'_i = P(\theta_i, \Phi(\vec{r}_i)) + Ch + \xi + \omega
\end{equation}
$Ch$ and $\xi$ represent the new features of rotation due to tumbling.
$Ch$ is a fixed part of the rotation and $\xi$ is a stochastic part,
chosen uniformly from an interval $[ -\eta, \eta]$ ($ \eta$
constant).  $\omega$ is an orientation term that takes, with equal
probabilities, one of the values 0 (forward direction) or $ \pi$
(backward direction). This orientation term gives the spinors their
name, as it make their re-orientation invariant to forward or backward
direction. $ \Phi(\vec{r}_i)$ is the local mean orientation in the
neighborhood of $ \vec{r}_i$. $P$ is a projection function that
represents the orientational interaction which acts on each spinor to
orient $ \theta_i$ along the direction $ \Phi(\vec{r}_i)$. $P$ is
defined by
\begin{equation}
P(\alpha, \beta) = \alpha + (\beta - \alpha) .
\end{equation}

Once oriented, the spinor advances a step $d$ in the direction
$ \theta'_i$, and the new location $ \vec{r'}_i$ is given by:
\begin{equation}
\vec{r'}_i = \vec{r}_i + d \left(\cos \theta'_i, \sin \theta'_i \right)
\end{equation}
The movement is confined within an envelope which is defined on the 
tridiagonal lattice. The step is not performed if $ \vec{r'}_i$ is outside 
the envelope. Whenever this is the case, a counter on the appropriate 
segment of the envelope is increased by one. When a segment counter 
reaches $ N_c$, the envelope advances one lattice step and a new 
lattice cell is occupied. Note that the spinor's direction is not reset 
upon hitting the envelope, thus it might "bang its head" against the 
envelope time and time again. The requirement of $ N_c $ hits 
represent the colony propagation through wetting of unoccupied areas 
by the bacteria. Note that $ N_c $ is related to the agar dryness, as 
more wetting fluid must be produced (more ``collisions'' are needed) to 
push the envelope on a harder substrate. 

Next we specify the mean orientation field $ \Phi$. To do so, we 
assume that each lattice cell (hexagonal unit area) is assigned one value 
of $ \Phi(\vec{r})$, representing the average orientation of the spinors 
in the local neighborhood of the center of the cell. The value of $ \Phi$ 
is set when a new lattice cell is first occupied by the advancement of 
the envelope, and then remains constant. We set the value of $ 
\Phi(\vec{r})$ to be equal to the average over the orientations of the $ 
N_c$ attempted steps that led to the occupation of the new lattice cell. 
The value of $ \Phi$ in any given point inside the colony is found by 
linear interpolation between the three neighboring centers of cells. 
Clearly, the model described above is a simplified picture of the 
bacterial movement. For example, a more realistic model will include 
an equation describing the time evolution of $\Phi$. However, the 
simplified model is sufficient to demonstrate the formation of chiral 
patterns. A more elaborate model will be presented elsewhere 
\cite{CGB99}.

Motivated by the presence of a maximal growth rate of the bacteria 
even for optimal conditions, each spinor in the model consumes food at 
a constant rate $\Omega_c $ if sufficient food is available. We 
represent the metabolic state of the $i$-th spinor by an 'internal energy' 
$E_i $. The rate of change of the internal energy is given by
\begin{equation}
{\frac{d E_i}{d t }} = \kappa C_{consumed} - {\frac{E_m 
}{\tau_R}}~,
\end{equation}
where $ \kappa $ is a conversion factor from food to internal energy ($ 
\kappa \cong 5\cdot 10^3 cal/g$) and $ E_m $ represent the total 
energy loss for all processes over the reproduction time $\tau_R$, 
excluding energy loss for cell division. $C_{consumed}$ is
$
C_{consumed} \equiv \min \left( \Omega_C , 
\Omega_C^{\prime}\right)~,
$
where $\Omega_C^{\prime}$ is the maximal rate of food consumption as
limited by the locally available food (Sec. \ref{sec:bio-bg}). When
sufficient food is available, $E_i$ increases until it reaches a
threshold energy. Upon reaching this threshold, the spinor divides
into two. When a 
spinor is starved for long interval of time, $E_i$ drops to zero and the 
spinor ``freezes''. This ``freezing'' represents entering a pre-spore
state (starting the process of sporulation, see section
\ref{sec:Cchemo}).

We represent the diffusion of nutrients by solving the diffusion 
equation for a single agent whose concentration is denoted by 
$n(\vec{x},t)$:
\begin{equation}
{\frac{\partial n}{\partial t}}= D_n \nabla^2C - b C_{consumed}~,
\end{equation}
where the last term includes the consumption of food by the spinors 
($b $ is their density). The equation is solved on the same tridiagonal 
lattice on which the envelope is defined. The length constant of the 
lattice $a_0$ must be larger than the size of the spinors' step $d$. The 
simulations are started with inoculum of spinors at the center and a 
uniform distribution of the nutrient. Both $\Phi$ and the spinors at the 
inoculum are given uniformly distributed random directions.

Results of the numerical simulations of the model are shown in Fig. 
\ref{fig:C-results}. These results do capture some important features 
of the observed patterns: the microscopic twist $Ch$ leads to a chiral 
morphology on the macroscopic level. The growth is via stable tips, all 
of which twist with the same handedness and emit side-branches. The 
dynamics of the side-branches emission in the time evolution of the 
model is similar to the observed dynamics.

For large noise strength $\eta$ the chiral nature of the pattern gives 
way to a branching pattern (Fig. \ref{fig:C-T}). This provides a 
plausible explanation for the branching patterns produced by \Cm 
grown on high peptone levels%  (Fig. \ref{fig:2.14})
, as the cells are 
shorter when grown on a rich substrate. The orientation interaction is 
weaker for shorter cells, hence the noise is stronger.

\section{The effect of chemotaxis}
\label{sec:chemo}

So far, we saw that the models can reproduce many aspects of the 
microscopic dynamics and the patterns in some range of nutrient level
and agar concentration, but at least for the \T-like growth, other
models can do the same \cite[and reference there in]{GKCB98}. We will
now extend the Non-Linear Diffusion model and the Communicating
Spinors model to test for their success in describing other aspects of
the bacterial colonies involving chemotaxis and chemotactic signaling
(which are believed to by used by the bacteria
\cite{BCC96,CCB96,BC97,BenJacob97}).

\subsection{Chemotaxis in the Non-Linear Diffusion Model}

As we mentioned in section \ref{sec:bg:chemotaxis},
in a continuous model we incorporate the effect of chemotaxis by 
introducing a chemotactic flux $ \vec{J}_{chem}$:
\begin{equation}
\vec{J}_{chem}\equiv \zeta (b )\chi (R)\nabla R
\end{equation}
$\chi (R)\nabla R$ is the gradient sensed by the bacteria (with $\chi
(R)$ having the units of 1 over chemical's concentration). $\chi (R)$
is usually taken to be either constant or the ``receptor law''. $\zeta
(b )$ is the bacterial response to the sensed gradient (having the
same units as a diffusion coefficient times the units of the bacterial
density $b$).  In the Non-Linear Diffusion model the bacterial
diffusion is $D_b=D_0b^k$, and the bacterial response to chemotaxis is
$\zeta (b)=\zeta _0 b \left( D_0 b^k \right) =\zeta _0 D_0 b^{k+1}$.
$\zeta _0$ is a constant, positive for attractive chemotaxis and
negative for repulsive chemotaxis.

We claimed that the fine radial branching patterns at low nutrient
levels result from repulsive chemotactic signaling. 
The equation describing the dynamics of the chemorepellent contains
terms for diffusion, production by pre-spores, decomposition by active
bacteria and spontaneous decomposition:
\begin{equation}
\frac{\partial R}{\partial t}= D_R{\nabla ^2 R}+s \Gamma _R-\Omega _R
b R- \lambda_R R
\label{r-eqn}
\end{equation}
where $D_R$ is a diffusion coefficient of the chemorepellent,
$\Gamma _R$ is an emission rate of repellent by pre-spores, $\Omega
_R$ is a decomposition rate of the repellent by active bacteria, and
$\lambda _R$ is a rate of self decomposition of the repellent.

Fig.\ \ref{fig:k-food-rep} demonstrates the effect of repulsive chemotactic 
signaling. In the presence of repulsive chemotaxis the pattern  becomes much 
denser with a smooth circular envelope, while the branches are thinner and 
radially oriented.

\subsection{Chemotaxis in the Communicating Spinors Model}
\label{sec:Cchemo}

The colonial patterns of \C morphotype (e.g. Fig. \ref{fig:C-results})
are rarely as ordered as the simulated patterns of the Communicating
Spinors model. For example, the branches of the observed colonies usually
have varying curvature.
In the simulations of \C morphotype shown in Fig. \ref{fig:C-results} all 
the branches have a uniform curvature. 
One of the reasons for this difference is the simplifications taken
during the model's development. A more elaborate model that we will
present \cite{CGB99} will be a better description of the colony.
However, some of the observed features can be explained in the
context of the Communicating Spinors model.
In some of the observed patterns (Fig. \ref{fig:2.16}b), the curvature
of the branches has a distinct relation to the branch's radial
orientation (the orientation relative to the radial direction):
the curvature is smaller when the branch is in the radial orientation
and larger when the branch is orthogonal to that orientation. This brings
to mind the radial organization of branches in the \T morphotype, and
indeed we where able to explain the chiral pattern with the aid of the
same concept -- repulsive chemotaxis.

Chemotaxis was introduced in previous versions of the Communicating 
Walkers model by varying, according to the chemical's gradient, the
probability of moving in different directions \cite{BSTCCV94a,BenJacob97}.
Modulating the directional probability is not the way bacteria
implement chemotaxis -- they modulate the length of runs. However, 
the growth of \T morphotype is insensitive to the details of the movement.
Modulating the directional probability is as good an
implementation of chemotaxis as many other implementations (it was
chosen for computational convenience). The pattern of the \Cm is based
on amplification of microscopic effects (singular perturbation) such
as the left bias in the bacterial tumbling. Small differences in
the microscopic dynamics of chemotaxis might affect the global
pattern. 
Indeed we found that modulating the directional probability yield 
unrealistic results in the simulations of \Cme. We had to resort to 
the bacterial implementation of chemotaxis -- modulating the length of 
runs according to the chemical's gradients.

When modulating the length of runs of walkers or spinors one must be 
careful not to change the particles' speed. Such change is not observed 
in experiments \cite{BP77,SBB86} and it has far reaching effects on 
the dynamics. Changing the particles' speed is like
changing the diffusion coefficient of the bacterial density field, 
a change that can have undesirable effects on the pattern.

Modulating the length of spinors' runs without changing their speed 
can be done by modulating the number of steps that compose a single 
run (that was our motivation for dividing the runs into steps). Since
the mean number of steps in a run is determined by the reorientation
probability $p$, chemotaxis should modulate this probability. For 
chemotaxis, the probability of changing direction by the $i$-spinor in 
one time step is (for a repellent $R$):
\begin{equation}
p^* = p + \chi(R) \partial_{\theta_i} R
\end{equation}
where $R$ is measured at the spinor position $\vec{r}_i$, $\chi(R)$ is 
the same as in the continuous model (either constant or the "receptor 
law") and $\partial_{\theta_i}$ is directional derivative in the spinor's 
direction $\theta_i$. $p^*$ is truncated to within the range $[0,1]$ as it
is a probability.
The length of the resulting runs will depend on the runs' direction, 
where a spinor moving up the gradient of the repellent will have 
shorter {\it mean} run length than the same spinor moving down the 
gradient.

The production and dynamics of the repulsive chemotactic signaling in the
Communicating Spinors model is the same as in the Non-Linear Diffusion
 model, see Eq. (\ref{r-eqn}) (with $s$ representing the density of
spinors that ``freezed'').
The patterns resulting from including repulsive chemotaxis in the
model have indeed branches with variable curvature, as can be seen in
Fig. \ref{fig:Cmodel:chemo}. The curvature is smaller for branches in
the radial direction. Food chemotaxis also varies the branches'
curvature, but in a less ordered manner, not similar to the observed
bacterial patterns.

Under different growth conditions the \Cm can produce very different 
patterns. As mention above, if the agar is soft enough the bacteria can 
move inside it . In such case, the bias in the bacterial movement might 
change or even reverse, and it is manifested in the curvature of the 
branches. Widely changing curvature of the branches can be seen in 
Fig. \ref{fig:2.16}a. The agar hardness was tuned such that in 
the beginning of the growth the bacteria could swim inside the agar, 
but they are forced to swim {\it on} the agar by the end of the
growth due to the marginal water evaporation during the growth.
In Fig. \ref{fig:Cmodel:long} we demonstrate the models' ability to
explain 
such patterns by changing the spinors' bias $Ch$ during the simulation. 
$Ch$ is set to be a continuous random function of the colonial size, 
which is constrained only at the beginning and end of growth to have
certain values. The function for $Ch$ is the same in all the images of
Figs. \ref{fig:Cmodel:long}, only that in various types of chemotaxis
are used.  As can be seen, repulsive chemotactic signaling is needed
to explain the observed bacterial patterns.

\section{Weak chirality in \Tm}

\label{sec:weak_chiral}

 Colonies of \Tm grown on hard substrate (above 2.0\% agar
concentration) exhibit branching patterns with a global twist with
the same handedness, as shown in Figs. \ref{fig:2.5} and
\ref{fig:weak}. Similar observations during growth of other bacterial
strains have been reported by Matsuyama \etal \cite{MM93,MHM93}. We
refer to such growth patterns as having weak chirality, as opposed to
the strong chirality exhibited by the \Cme.

In \cite{BCSCV95}, Ben-Jacob \etal proposed that, in the case of \Tme,
it is the high viscosity of the
"lubrication" fluid during growth on a hard surface that replaces the
cell-cell co-alignment of the \Cm that limit the rotation of tumbling.
They further assumed that the rotation should be relative to a specified
direction.
They used gradient of a chemotaxis signaling field (specifically, the
long-range repellent
chemotaxis) as a specific direction, rather than the local mean
orientation field which is used in the case of \Cme.
It was shown in \cite{BCSCV95} that inclusion of the above features
in the Communicating Walkers model indeed leads
to a weak chirality which is highly reminiscent of the observed one. The
idea above also provides a plausible explanation to the observations of
weak chirality by Matsuyama \etal
\cite{MM93}
in strains defective in production of "lubrication" fluid.

\subsection{Weak chirality -- the Non-Linear Diffusion Model}

In the reaction-diffusion model, weak chirality can obtained by
modifying the chemotactic mechanism
and causing it to twist: We alter the expression for the
chemotactic flux $\vec{J}_{chem}$ (Eq. \ref{j_chem}) so that it is
not oriented with the chemical gradient ($\nabla R$) anymore. Instead
it is oriented with a rotated vector ${\bf \hat{R} (\theta)}\nabla
R$, where ${\bf \hat{R} (\theta)}$ is the two-dimensional rotation
operator and $\theta$ is the rotation angle.
The chemotactic flux is thus written:
\begin{equation}
\vec{J}_{chem} = \zeta (b )\chi (R)({\bf \hat{R} (\theta)}\nabla R)
\label{j_chem_chiral}
\end{equation}
The effect of rotating the repulsive chemotaxis, as depicted in
Fig. \ref{fig:k-weak}, is to make the pattern chiral, with the degree
of chirality determined by the rotation angle $\theta$.

One must note that adding a similar rotation to the food chemotaxis does
not have the same effect, because the nutrient gradients do not posses
the long-range, radial nature of the chemorepellent gradients.

\subsection{Weak chirality -- the Communicating Spinors Model}

As was demonstrated in section \ref{sec:C}, the
Communicating Spinors model is robust enough to reproduce patterns of \Tme,
as well as patterns of \Cme.
Here we use it to model patterns of weak chirality. 
Two simulated \T-like colonies are shown in Fig.
\ref{fig:Cmodel:weak}. Fig. \ref{fig:Cmodel:weak}a shows colony with
radial branches while Fig. \ref{fig:Cmodel:weak}b shows colony with
weak chirality and thinner branches. In the two simulations the
spinors have exactly the same response to chemotaxis and the same bias
$Ch=9^\circ$. The two runs differ in the freedom of rotation $\eta$; in
\ref{fig:Cmodel:weak}a the spinors can rotate freely ($\eta=180^\circ$)
while in \ref{fig:Cmodel:weak}b the spinors rotation is somewhat
limited ($\eta=35^\circ$, while for the colony of strong chirality
$\eta=5^\circ$). 

It seems that both models -- the Non-Linear Diffusion model and the
Communicating Spinors model -- can capture the essential features of
the observed weak chirality. Yet a closer examination reveals that the
description of the two models is incompatible. In the Non-Linear
Diffusion model the bias from the direction of the gradient is
through the chemotaxis process. The spinors, like the bacteria, cannot
modulate their runs as a function of the difference between their
direction and that of the gradient; they do not know what is the
direction of the gradient, only the directional derivative along their
path.
As was demonstrated in figure \ref{fig:Cmodel:weak}, one of the key
features for the weak chirality in the spinors model is the
correlation in orientation of the spinors (through $\Phi$). 
In fact in this model the twist of the branches stems from the
deviation of the runs' direction from the orientation of neighboring
spinors. The twist of the branches is related only indirectly, through
the neighbors' orientation, to the chemorepellent's gradient.
A continuous model of such process should include information about the
mean orientation of the bacterial cells. It should include chemotaxis
without rotation, anisotropic diffusion (smaller diffusion coefficient
in orientations orthogonal to the mean orientation of neighbors) and a
rotation on the diffusion operator. Such a model will be presented
elsewhere \cite{CGB99}.

The discrete spinors model allows for a detailed representation of the
bacterial properties.
The macroscopic dynamics and resulting patterns, however, are similar
in both models -- apparently the growth does not amplify the
difference in the microscopic dynamics.
Thus the `unrealistic' microscopic description of the Non-Linear
Diffusion model does not rule it out as an approximation to the growth
dynamics of bacterial colonies with weak chirality.

\subsection{Chirality measure}

\label{sec:measure}

All through this manuscript we referred to chirality as a two-valued
property -- either the pattern is chiral or it is not. There are various
attempts in the literature to quantify chirality with a continuous
measure.  See for example the method of Avnir \etal \cite{ZA95,KHA96},
who
applied their method also to large disordered objects.
While this method is general and can quantify with a
single number the measure of asymmetry of any given object, it is not
quite satisfactory for our purpose. We would like to know the time
evolution of the chirality of a colony, and not just ``mean''
chirality given by a single number. We sacrifice the generality of the
measure to that end.

Since the growth velocity of the colonies (both experimental and
simulated) is constant, we measure the chirality as a function of
radius instead of function of time. Thus we can work on chirality of
an image, not of a process. The image can be a scanned picture of the
real colony or a result of a computer simulation.
We look for a mapping of the image to a new one, which in
some sense does not distinguish left from right (the ambiguity stems
from the fact that a large random object will not have, in general,
reflection symmetry, thus there is no trivial definition for chirality
of such objects).
The mapping is defined by:
\begin{equation}
(r,\theta) \rightarrow (r,\theta + \Delta \theta (r))
\end{equation}
where each point in the image is described by the polar coordinates
$(r,\theta)$, measured from the center of the colony.
Thus, each point is rotated by an increment $\Delta \theta$ which
depends on the radius $r$ (i.e. the distance from the center).

Working on many experimental patterns, as well as simulated patterns,
we have learned that in most cases a linear dependence of $\Delta
\theta$ on $r$ is sufficient to give quite satisfactory results, that
is, to transform a chiral pattern into a ``normal'' branching pattern.
The rotating angle is thus written:
\begin{equation}
\Delta \theta (r) =  \left(\frac{r}{r_{max}}\right)\theta_{max}
\end{equation}
where $r_{max}$ is the radius of the colony, and $\theta_{max}$ is the
rotation angle at that radius.

The fact that this linear angular mapping suffices to ``de-chiral''
the simulated patterns may not be of much importance (in the case of
the continuous model, at least, this is almost a direct result of the
way in which we introduce the weak chirality). The same
transformation works for images of real colonies of \Tme, but {\it
does not} work for chiral colonies of other bacteria (see Sec.
\ref{sec:conc}).  This strengthens our belief in the models.

\section{Conclusions}
\label{sec:conc}

We first briefly reviewed experimental observations of colonial
patterns formed by bacteria of the species \dend*.
We described both the tip-splitting growth of the \Tm and the chiral,
twisted-branches growth of \Cm.
Both colonial patterns and
optical microscope observations of the bacteria dynamics were
presented.

In this manuscript we presented observations of various forms of
chiral patterns in bacterial colonies. Our goal was to explain the
various aspects of chirality.
We used two types of models: continuous
reaction-diffusion models which deal with bacterial density, and a
hybrid semi-discrete model which deals with properties of the
individual bacterium.
From a comparison of the models' simulation and experimental
observations we conclude that chemotactic signaling plays an
important part in the development of colonies of the two types.
We also estimate how sensitive the growth is to the details of the
microscopic dynamics, demonstrating that the more 'complex' the
pattern is, the more sensitive the growth is to the small details.

We would like to note that the \dend is not the only bacteria whose
colonies exhibit chirality. Ben-Jacob \etal discussed in
\cite{BCCVG97} the formation of colonies of \Vname*, where each branch
is produced by a leading droplet and emits side branches, each with
its own leading droplet.  Each leading droplet consists of hundreds to
millions of bacterial cells that circle a common center (a vortex) at
a cellular speed of about $10\mu$m/s.
In Fig. \ref{fig:V} we show a colonial pattern of these bacteria. The
chirality we termed `weak chirality' is evident in this figure. In
this case the chirality is not related to the handedness of the
flagella, but to the rotation of the vortices. When ``pushed'' by
repulsive chemotaxis, Magnus force acts on the vortices and drive
them side-ways from the radial direction of the chemorepellent's
gradient.
This difference in mechanisms is expressed in the global pattern:
the colonial patterns of \Vname cannot be ``de-chiraled'' by the
transformation (linear angular mapping) that ``de-chiral'' the \Tme.
The fact that the models for weak chirality match in this respect the
the weak chirality of \Tm and not the `weak chirality' of \Vname is
another support for their success in describing the bacterial
colonies.

We hope we have convinced the reader that chirality in patterns of
bacterial colonies gives important clues about the underlying
dynamics. The processes leading to such patterns are more complex than
those leading to non-twisted branching patterns. The chiral patterns
are more sensitive to the underlying dynamics and as such they require
more accurate models.  This reflects on the success of the models we
presented as being a good description of the colonies.

\section*{Acknowledgments}
Identifications of the \dend* and genetic studies are carried in
collaboration with the group of D.\ Gutnick. Presented studies are
supported in part by a grant from the Israeli Academy of Sciences
grant no.\ 593/95, by the Israeli-US Binational Science Foundation BSF
grant no. 00410-95 and by a grnat from IMK Almene Fond.
Two of us, E.\ Ben-Jacob and I.\ Golding, thank the IMA for
hospitality during part of this project.
One of us, I. Cohen, thanks The Colton Scholarships for their support.

\bibliographystyle{plain}

\section*{Figure Captions}

\begin{enumerate}
% \begin{figure}[htbp]
\item { % \caption {
\label{fig:intro}
(a) Typical example of branching growth of the \Tname* (referred to as
\Tm \protect\cite{BCG98})for 0.5 g/l peptone
and 1.75 \% agar concentration.
(b) Chiral growth of the \Cname* (referred to as
\Cm \protect\cite{BCG98}) for 2.5 g/l peptone and 1.25 \% agar concentration.
}
% \end{figure}

% \begin{figure}[htbp]
\item { % \caption {
\label{fig:2.1}
Patterns exhibited by the \T morphotype as function of peptone
level (increasing from left to right) and agar concentration (1.5\%
bottom row, 2\% middle row, 2.5\% top row).
}
% \end{figure}

% \begin{figure}[htbp]
\item { % \caption {
\label{fig:2.2}
Examples of typical  patterns of \Tm for intermediate agar
concentration.
(a) Compact growth for 12$g/l$ peptone level and 1.75\% agar
concentration.
(b) Dense fingers for $3g/l$ peptone and 2\% agar.
(c) Branching fractal pattern for $1g/l$ peptone and 1.75\% agar.
(d) A pattern of fine radial branches for $0.1g/l$ peptone and 1.75\%
agar.
}
% \end{figure}

% \begin{figure}[htbp]
\item { % \caption {
\label{fig:2.3}
Colonial patterns of \Tm.
(a) Fractal pattern for $0.01g/l$ peptone level and 1.75\%
agar concentration.
(b) Dense branching pattern for $4g/l$ peptone and 2.5\% agar.
Note that the branches are much thinner than those in Fig.
{\protect\ref{fig:2.2}}b, i.e.
the branches are thinner for higher higher agar concentrations.
}
% \end{figure}

% \begin{figure}[htbp]
\item { % \caption {
\label{fig:2.4}
(a) Pattern of concentric rings superimposed on a branched colony for
2.5$g/l$ peptone level and 2.5\% agar concentration.
(b) Concentric rings in a compact growth for $15g/l$ peptone level and
2.25\% agar concentration.
}
% \end{figure}

% \begin{figure}[htbp]
\item { % \caption {
\label{fig:2.5}
Weak chirality exhibited by the \Tm during growth on 4$g/l$ peptone and
2.5\% agar concentration.
}
% \end{figure}

% \begin{figure}[htbp]
\item { % \caption {
\label{fig:2.6}
Density variations within branches of a colony of \Tm. Optical
microscope observations $\times50$.
}
% \end{figure}

% \begin{figure}[htbp]
\item { % \caption {
\label{fig:2.9}
Closer look on branches of a colony.
a) $\times20$ magnification shows the sharp boundaries of the
branches.
The width of the boundary is in the order of micron.
b) Numarsky (polarized light) microscopy shows the hight of the
branches and their envelope. What is actually seen is the layer of
lubrication fluid, not the bacteria.
c) $\times50$ magnification shows the bacteria inside a branch. Each
bar is a single bacterium. There are no bacteria outside the branch.
}
% \end{figure}

% \begin{figure}[htbp]
\item { % \caption {
\label{fig:2.10}
Electron microscope observation of \T bacteria. Round or oval shapes
with bright center are spores. Elongated shapes are living cells.
The cells engulfing oval shapes are pre-spores.
}
% \end{figure}

% \begin{figure}[htbp]
\item { % \caption {
\label{fig:2.14}
Patterns exhibited by the \Cm for different growth conditions.
a) Thin disordered twisted branches at 0.5$g/l$ peptone level and
1.5\% agar concentration.
b) Thin branches, all twisted with the same handedness. at 2$g/l$
peptone level and 1.25\% agar concentration.
c) Pattern similar to (b) but on softer agar: 1.4 $g/l$ peptone level
and 0.75\% agar concentration.
d) Four inocula on the same plate, conditions of 1$g/l$ peptone level
and 1.25\% agar concentration.
}
% \end{figure}

% \begin{figure}[htbp]
\item { % \caption {
\label{fig:2.16}
In agar soft enough for the bacteria to swim in it, the branches lose
the one-side handedness they have on harder agar. The two colonies of
(a) and (b) are of 5$g/l$ peptone level and 0.6\% agar concentration.
The two patterns are of two stains of the \Cme, strains whose patterns
are indistinguishable on harder agar.  c) Closer look ($\times 10$
magnification) on a colony grown at 8$g/l$ peptone level and 0.6\%
agar concentration.
}
% \end{figure}

% \begin{figure}[htbp]
\item { % \caption {
\label{fig:2.17}
Optical microscope observations of branches of \Cm colony.
a) $\times20$ magnification of a colony at 1.6$g/l$ peptone level and
0.75\% agar concentration, the anti-clockwise twist of the thin
branches is apparent. The curvature of the branches is almost constant
throughout the growth.
b) $\times10$ magnification of a colony at 4$g/l$ peptone level and
0.6\% agar concentration at branches are not thin, but have a feathery
structure. The curvature of the branches varies, but it seems that at
any given stage of growth the curvature is similar in all branches.
That is, the curvature is a function of colonial growth.
c) $\times500$ magnification of a colony at 1.6$g/l$ peptone level and
0.75\% agar concentration. Each line is a bacterium. the bacteria are
long (5-50$\mu m$) and mostly ordered.
}
% \end{figure}

% \begin{figure}[htbp]
\item { % \caption {
Effect of varying the initial nutrient concentration $n_0$ on colony 
pattern. $n_0$ increases from left to right:
1.2 (a), 1.4 (b), 1.7 (c), 2 (d), 3 (e), 6 (f).
The minimal value of $n_0$ to support growth is $1$.
}
\label{fig-nutrient}
% \end{figure}

% \begin{figure}[htbp]
\item { % \caption {
Effect of varying $\lambda$, the fluid absorption rate, on colony pattern. 
The fluid production rate $\Gamma$ is $1$ in the upper row
and $0.3$ in the lower row.
In both rows $\lambda$ increases from left to right:
$\lambda=0.03$ (left), $\lambda=0.1$ (center), $\lambda=1$ (right)
The patterns become more ramified as $\lambda$ increases. 
Decreasing $\Gamma$ also
produces a more ramified pattern.
The other parameters are:
$D_b=D_l=1,  n_0=1.5$
}
\label{fig-lambda}
% \end{figure}

% \begin{figure}[htbp]
\item { % \caption {
\label{fig:k1}
Growth patterns of the Non-Linear Diffusion model, for different
values of initial nutrient level $n_0$. Parameters are: $D_0=0.1,
k=1, \mu=0.15$. The apparent 6-fold symmetry is due to the underlying
tridiagonal lattice.
}
% \end{figure}

% \begin{figure}[htbp]
\item { % \caption {
\label{fig:C-results}
A morphology diagram of the Communicating Spinors model for various
values of $N_c$ and initial $n$ concentration $n_0$.
$Ch=6^\circ$, $\eta=3^\circ$, $d=0.2$, $p=0.5$.
}
% \end{figure}

% \begin{figure}[htbp]
\item { % \caption {
\label{fig:C-T}
When the noise $\eta$ is increases to $\eta=180^\circ$ the tumbling of
the spinors becomes unrestricted. Their movement becomes like that of
the \T bacteria and accordingly the simulated colonial pattern is like
that of \Tme.
On the left $\eta=3^\circ$, on the right $\eta=180^\circ$.
}
% \end{figure}

% \begin{figure}[htbp]
\item { % \caption {
\label{fig:k-food-rep}
Growth patterns of the Non-Linear Diffusion model with food
chemotaxis (left, see section \protect\ref{sec:bio-bg}) and repulsive chemotactic signaling (right)
included. 
$\chi _{0f}=3,\chi _{0R}=1,D_R=1,\Gamma _R=0.25,\Omega _R=0,\Lambda
_R=0.001$. Other parameters are the same as in figure
\protect\ref{fig:k1}.
The apparent 6-fold symmetry is due to the underlying tridiagonal
lattice.
}
% \end{figure}

% \begin{figure}[htbp]
\item { % \caption {
\label{fig:Cmodel:chemo}
The effect of repulsive chemotactic signaling on the Communicating
Spinors model.
a) Without chemotaxis.
b) With repulsive chemotaxis.
The Spinors are repelled from the inner parts of the
colony. The resulting curvature of the branches is reduces when they
are in the radial direction.
In spite the reversed handedness, the pattern resemble Fig.
\protect\ref{fig:2.16}b.
}
% \end{figure}

% \begin{figure}[htbp]
\item { % \caption {
\label{fig:Cmodel:long}
The snake-like branches observed in Fig. \protect\ref{fig:2.16}a can be
reproduced by the Communicating Spinors model. $Ch$ is a continuous
function of the colony's radius (the same function in a, b, and c).
Maximal value of $Ch$ is $8^\circ$, minimal value is $-2^\circ$.
(a) With repulsive chemotactic signaling.
(b) Without chemotaxis.
(c) With food chemotaxis.
The best resemblance to the observed colony is obtained with
repulsive chemotactic signaling.
}
% \end{figure}

% \begin{figure}[htbp]
\item { % \caption {
\label{fig:weak}
Weak chirality (global twist of the branches) exhibited by the \Tme
for a peptone level of 0.25$g/l$ peptone level and agar concentration
of 1.75 \%. 
}
% \end{figure}

% \begin{figure}[htbp]
\item { % \caption {
\label{fig:k-weak}
Growth patterns of the Non-Linear Diffusion model with a ``squinting''
repulsive
chemotactic signaling, leading to weak chirality.
Parameters are as in the previous picture, $\theta= 43^o$.
}
% \end{figure}

% \begin{figure}[htbp]
\item { % \caption {
\label{fig:Cmodel:weak}
Weak chirality of the \Tm is modeled by the Communicating Spinors
model. Both simulations are with repulsive chemotactic signaling and
with bias in the walkers rotation. 
a) With free rotation ($\eta=180^\circ$) The pattern is branched,
without apparent chirality.
expressed.
b) With constrained rotation ($\eta=35^\circ$) weak chirality is
expressed.
}
% \end{figure}

% \begin{figure}[htbp]
\item { % \caption {
\label{fig:V}
A colony of \Vname* on 10 $g/l$ peptone level and 2\% agar
concentration. The dots at the tips of the branches are bacterial
vortices -- each is composed of up to millions of bacterial cells
rotating around a common center. The twist of the branches results
from a Magnus force induced by repulsive chemotactic signaling.
}
% \end{figure}
\end{enumerate}

\end{document}